\documentclass[twocolumn, showpacs, prb]{revtex4-1}

\def\ind#1{{_{\mathrm{#1}}}}

\usepackage{amssymb}
\usepackage{amsmath}
\usepackage{graphicx,color}

\begin{document}
\title{Entanglement between nitrogen vacancy spins in diamond  controlled by a nano mechanical resonator}

\author{L. Chotorlishvili$^1$, D. Sander$^2$, A. Sukhov$^1$, V. Dugaev$^{1,3}$,
V. R. Vieira$^4$, A. Komnik$^5$, J.  Berakdar$^1$}

\address{$^1$Institut f\"ur Physik, Martin-Luther Universit\"at Halle-Wittenberg, D-06120 Halle/Saale, Germany \\
$^2$Max Planck Institute of Microstructure Physics, D-06120 Halle/Saale, Germany\\
$^3$Department of Physics, Rzesz\'{o}w University of Technology Al. Powstanc\'{o}w Warszawy 6, 35-959 Rzesz\'{o}w, Poland\\
$^4$Department of Physics and CFIF, Instituto Superior T\'{e}cnico, Universidade T\'{e}cnica de Lisboa, Avenida Rovisco Pais, 1049-001 Lisboa, Portugal\\
$^5$Institut f\"ur Theoretische Physik, Universit\"at Heidelberg, Philosophenweg 19, D-69120 
}
\date{\today}

\begin{abstract}
We suggest  a new type of nano-electromechanical resonator, the functionality of
which is based on a magnetic field induced deflection of an appropriate cantilever that
oscillates between nitrogen vacancy (NV)  spins in daimond.
 Specifically, we consider a $\rm{Si(100)}$
cantilever coated with a thin magnetic $\rm{Ni}$ film. Magnetoelastic stress
and magnetic-field induced torque are utilized to induce a controlled cantilever
deflection. It is shown that, depending on the value of the system parameters,
the induced asymmetry of the cantilever deflection substantially modifies the
characteristics of the system. In particular, the coupling strength  between the
NV spins and the degree of entanglement can be controlled through magnetoelastic
stress and magnetic-field induced torque effects. Our theoretical proposal can
be implemented experimentally with the potential of  increasing several times the coupling
strength between the NV spins as  compared to the maximal coupling strength
reported before in P. Rabl, \emph{et al.}
Phys. Rev. B {\bf 79}, 041302(R) (2009).
\end{abstract}

\pacs{}
\maketitle
\section{Introduction}
Nano-electromechanical   resonators (NEMs) are attracting intense research efforts
 due to a number of favorable properties such as the high sensitivity  and the swift response
 to an external  force with a low power consumption. This makes NEMs
 attractive for applications, e.g. for microwave switches, nano-mechanical memory elements, and for single molecule sensing \cite{Naik09,Connell,Safa}.
 The role
of quantized  mechanical motion coupled to other quantum degrees of freedom as well as the influence of dissipation and noise
  are  important issues in NEMs research with numerous
   findings and demonstrations of applications, e.g.
 Refs.~\cite{qm1,qm2,qm3,qm4,qm5,qm6,qm7,qm8,MeMc11,AtIs11,Rabl10,Pran11,LuHa10,ScBo10,KaCr09,ChUg11,ShOm12,LiMi10,ShAs10,ZuRe09} and
 further references therein.
  %
These studies also evidence the potential of  NEMs  for studying fundamental questions concerning the
 quantum-classical  interrelation and issues related to entanglement and quantum correlations.
Particularly interesting for the  present work are spin states in
nitrogen vacancy (NV) impurities in
diamond\cite{RaCa09,ZhWe10,ArJa11,RuBu04,Treu12,KoBl12,WaDo11,WrJe06} as utilized for
quantum information studies. Their quantum mechanical properties can be mapped onto
effective two-level systems which possess very long decoherence times even at a
room temperature. On the other hand, they allow for a high degree of tunability
via external magnetic fields which renders possible the use of NV impurity
spins as sensors. For instance, a detection of a single
electronic spin by a classical cantilever was demonstrated in
\cite{RuBu04}.  Magnetic tips attached on the free end of a cantilever
generate a magnetic field gradient during  oscillations, which  induces a magnetic coupling between the nanomechanical oscillator and the spin system. Using the backaction due to this coupling one can readout the cantilever motion \cite{RaCa09,ZhWe10}.\\
One further fascinating application for the NV spin based nano-electromechanical resonator system is magnetic resonance force microscopy (MRFM) which was proposed to improve the detection resolution for the three-dimensional imaging of macromolecules\cite{RuBu04}. Operation of the MRFM is based on the detection of the magnetic force between a ferromagnetic tip and spins. The possibility to achieve strong (up to $0.1~$MHz or even stronger) coupling between the quantized motion of the nano-resonator and NV impurity spin was demonstrated in Ref.~\cite{KoBl12}. However, an efficient control of the coupling strength and the development of ultrasensitive cantilever-based force sensors is still a fundamental challenge. Yet another application is based on using such a resonator in order to produce controllable entanglement between spins. Despite rather long decoherence time scales it has not been demonstrated yet, to our knowledge.

In this paper we propose a new type of a nano-electromechanical resonator system,
the functionality of which is based on the NV spin controlled by a magnetic
field. The nano-resonator is a cantilever, which is covered with a magnetic
film, such that we can exploit magnetoelastic stress or magnetic torque effects for a controllable
cantilever deflection.
At the free end of the cantilever magnetic tips are mounted, as indicated
in Fig.~1. The magnetic field allows for a full control of the cantilever
configuration via the amplitude of the applied magnetic field. This is the
major novelty of the current proposal. As will be demonstrated below, the field-induced
deflection of the cantilever has a major impact on the interaction strength
between the NV spins and the magnetic tips, as well as on the strength of the
indirect interaction between the spins mediated by the cantilever.
We will show that the asymmetry of the nanomechanical system with respect to the shape of the cantilever controlled by magnetic fields drastically changes the degree of entanglement.

The paper is organized as follows. In the next section we introduce our system as well as the model Hamiltonian and discuss the relevant experimental parameters. Section \ref{deflection} describes the magnetic field induced cantilever deflection. The subsequent Section \ref{interaction} focuses on the effects of the induced indirect interaction of the impurities mediated by the cantilever coupling. The influence of the asymmetry coupling on the degree of entanglement is considered in Section \ref{asymmetry}. Finally, Section \ref{results} summarizes the results gives
some perspectives for further progress.

\begin{figure}[h]
\centering \includegraphics[scale=.5]{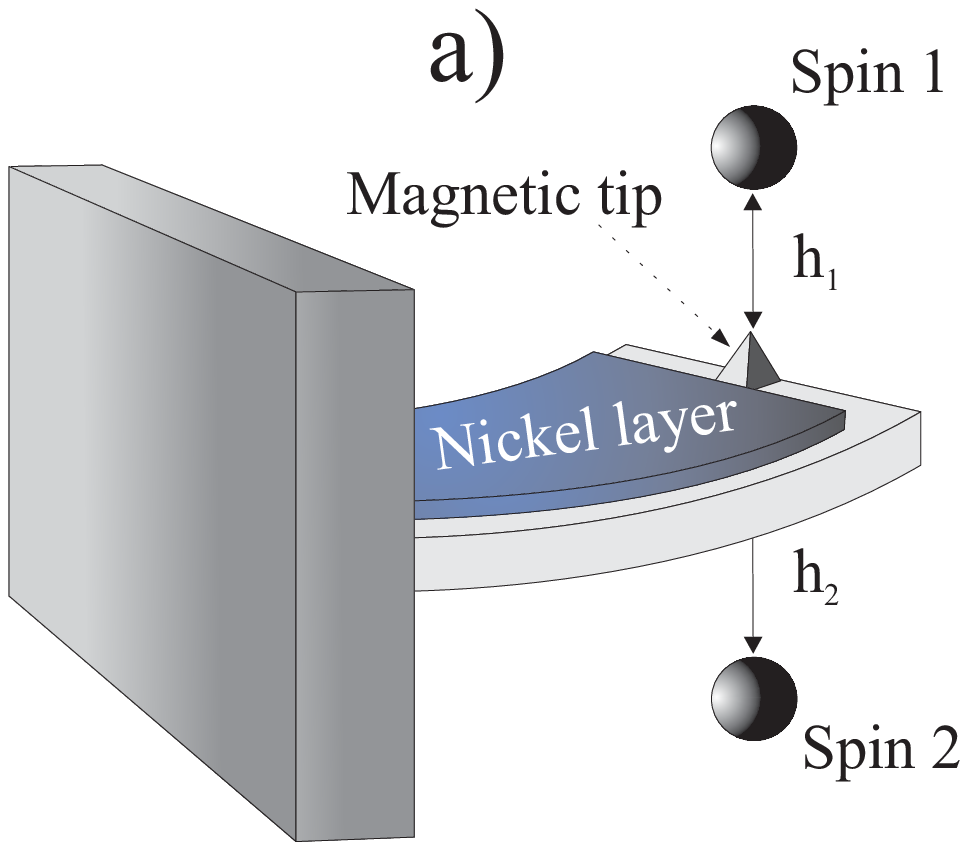}
\centering \includegraphics[scale=.5]{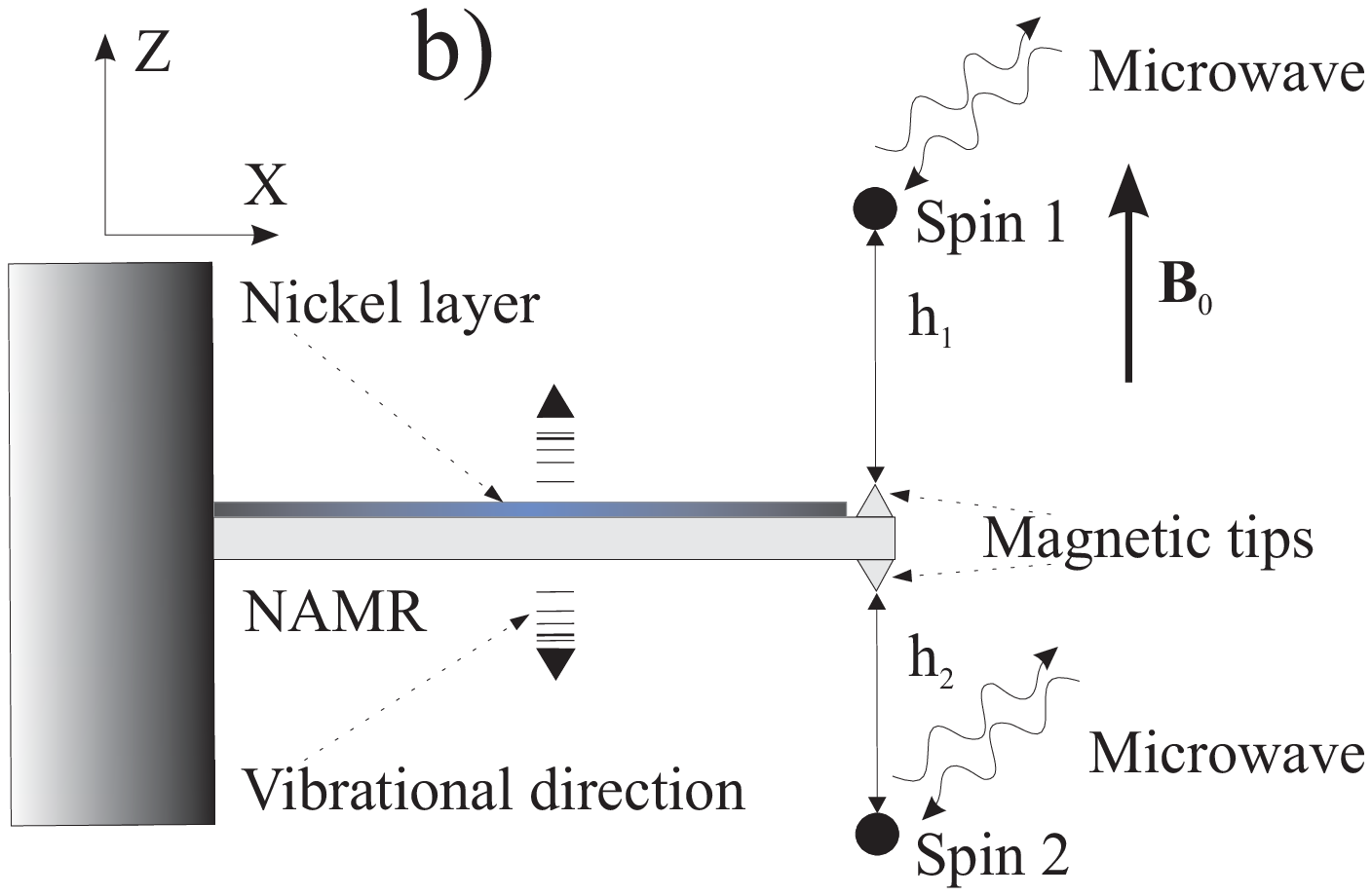}
\caption{Schematics of the nanomechanical resonator system discussed in the present project.
a) The distances $h_{1,2}$ between the NV spins and magnetic tips  can be adjusted by the field-induced deflection of the cantilever. The deflection is controlled by the applied magnetic field $B_{0}$. In the absence of the external constant magnetic field
$B_{0}=0$ the system becomes symmetric, and $h_{1}=h_{2}=h_{0}\approx 25$~nm b).
Microwaves are used to drive spin transitions at the NV centers of spin 1 and spin 2.
b) Two magnetic tips are mounted on the free end of the cantilever, which mechanically oscillates in the z-axis direction.
The length of the cantilever is of the order $L\sim 3000$~nm . The cantilever surface is covered with a $10$~nm thin Ni film. See Section \ref{deflection} for further details.}
\label{fig_1}
\end{figure}

\section{Theoretical model}
\label{theory}

\begin{figure}[h]
\centering \includegraphics[scale=.85]{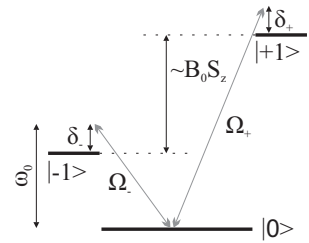}
\caption{Configuration of the energy levels. $\Omega_{\pm}$ stand for the Rabi  frequencies between the ground
and the excited levels $(\mid0>,\mid-1>)$ and $(\mid0>,\mid1>)$. $\delta_{\pm}$
denote the detuning between microwave frequency $\omega_{0}$ and transition frequencies.
}
\label{fig_2}
\end{figure}
The system of our interest is schematically shown in Fig. \ref{fig_1}.
Magnetic tips are attached to the free ends  of the cantilever on both sides.
The deflection of the cantilever and consequently the distance between magnetic
tips and spins $h_{1,2}$ is controlled via the magnetic field $\vec{B}_0$
applied along the $Z$ axis. The nitrogen vacancy center in diamond consists of a
substitutional nitrogen atom with an adjacent vacancy. The total spin of the
many-electron orbital ground state of the NV center is described by the spin
triplet $S=1$, $m\ind{S}=-1, 0, 1$. States with different $|m\ind{S}|$ are
separated by a zero field splitting barrier \cite{RaCa09} which is of the order
$\omega_0=2.88$~GHz. This  splitting is an intrinsic property of the NV spin system
and originates \cite{WrJe06} from the effect of the spin-orbit and spin-spin
interactions leading to the single-axis spin anisotropy
$DS^2_{\mathrm{z}}\approx\hbar\omega_{0}$. In the theoretical description we set
$\hbar=1$. The role of the applied external magnetic field is twofold:
First of all, the applied external magnetic field $\mu\ind{B}B_0<\hbar\omega_0$,
due to the Zeeman shift proportional to $B_0S\ind{z}$,
removes the degeneracy
of the levels $|-1>, |1>$.
Besides, as will be shown below, the external magnetic field, due to the
thin magnetic $\rm{Ni}$ film deposited on the cantilever,
modifies the shape of the cantilever. We will
demonstrate that the asymmetry of the cantilever position between spins 1 and 2 (Fig.~1), induced by an external magnetic field,
has important consequences for the coupling strength and the
entanglement between the NV spins. Therefore, one can control the degree of entanglement and increase the interaction strength
between the NV spins. This constitutes  an important advance in nanomechanics.

The Hamiltonian of the single NV spin system reads \cite{RaCa09}
\begin{equation}
\displaystyle H\ind{NV}  = \sum\limits_{i =  \pm 1} {\left( { - \delta _i \left| i \right\rangle \left\langle i \right| + \frac{{\Omega _i }}
{2}\left( {\left| 0 \right\rangle \left\langle i \right| + \left| i \right\rangle \left\langle 0 \right|} \right)} \right)}.
\label{eq_1}
\end{equation}
In the case of a weak magnetic field $\mu\ind{B}B_0\ll \hbar\omega_0$  ($\ll30$~mT) one can neglect
level splitting and set $\delta_-=\delta_+,~~ \Omega_-=\Omega_+$. In this case
the Hamiltonian (\ref{eq_1}) couples the ground state $|0>$ to the ``bright''
superposition of the excited states $\displaystyle
|b>=\frac{1}{\sqrt{2}}\big(|-1>+|1>\big)$, while the ``dark'' state $\displaystyle
|d>=\frac{1}{\sqrt{2}}\big(|-1>-|1>\big)$ is readout and decoupled from the
process. Since only two states are involved, the NV spin triplet can be
described via a $S=1/2$ pseudo-spin model. If the external constant magnetic
field is strong enough, the splitting between the levels $|-1>, |1>$ is larger
and the system becomes identical to the three level generalized Jaynes-Cummings model in the so called lambda configuration\cite{ChTo08,YoEb85} (See Fig. \ref{fig_2}).\\
A peculiarity of the generalized Jaynes-Cummings model is the expectation of two different
transition frequencies between states $(|0>,|-1>)$ and $(|0>,|1>)$. The
Hamiltonian of the system holds a $SU(3)$ symmetry and can be specified in terms
of the Gell-Mann generators\cite{ChTo08}. The three-level Jaynes-Cummings model
is exactly solvable in the general case. If the rf-field contains only one frequency
resonant to the transition $(|0>,|-1>)$ the system can be reduced to the $S=1/2$
pseudo-spin model which will describe transitions between the states $|-1>, |0>$
only, since the transition between the levels $|1>$, $|0>$ is off resonance and
therefore forbidden. However, the Rabi frequency $\Omega_-$ of the transition
between the states $|-1>, |0>$ in this  case is different from the transition
frequency between the ground state $|0>$ and the bright superposition of the
excited states $\displaystyle |b>=\frac{1}{\sqrt{2}}\big(|-1>+|1>\big)$.
Nevertheless, in both cases our system can be described via the effective
two-level model with a different Rabi transition frequency.
The Hamiltonian of the system in the frame rotating with the frequency of the rf-field has the form
\begin{equation}
\displaystyle H\ind{S}  =  - \delta \left| { - 1} \right\rangle \left\langle { - 1} \right| + \frac{{\Omega \left( B \right)}}
{2}\left( {\left| { - 1} \right\rangle \left\langle 0 \right| + \left| 0 \right\rangle \left\langle { - 1} \right|} \right).
\label{eq_2}
\end{equation}
Here $\delta=\omega_0-\Omega(B)$ is the detuning between the microwave frequency
and the intrinsic frequency of the spins. The Rabi frequency of the transition
between up and down spin states $\Omega(B)$ depends on the amplitude of the
magnetic field and, thus, includes both limiting cases. In the case of zero external
constant magnetic field $B=0$ and zero splitting between the levels $|-1>, |1>$
the Rabi frequency for the transition between the bright $|b>$ and the ground $|0>$
states is equal to $\Omega(0)=\Omega_0$, while in the case of a nonzero external
field and nonzero splitting for the transition between levels $|-1>, |0>$ the Rabi frequency is equal to:
\begin{equation}
\Omega \left( B \right) = \Omega _0  - \Delta \Omega \left( B \right), \,\,\,\, \Delta \Omega \left( B \right) = \mu\ind{B} \left( {B_0  + B\ind{ms}} \right).
\label{eq_3}
\end{equation}
Here $B_0$ is the amplitude of the external constant magnetic field applied on
the system, $B\ind{ms}$ is the magnetic field produced by the magnetic film on
the cantilever at the position of the spins Spin 1 and 2, and it is $B_0>B\ind{ms}$. The
eigenbasis of the Hamiltonian (\ref{eq_2}) is given by the following states\cite{ZhWe10}
\begin{eqnarray}
\label{eq_4}
  &&\left| g \right\rangle  = \cos \left( {\theta /2} \right)\left| { - 1} \right\rangle  + \sin \left( {\theta /2} \right)\left| 0 \right\rangle , \nonumber \\
  &&\left| e \right\rangle  =  - \sin \left( {\theta /2} \right)\left| { - 1} \right\rangle  + \cos \left( {\theta /2} \right)\left| 0 \right\rangle , \hfill \\
  &&\tan \theta  =  - \frac{\Omega }{\delta } \nonumber.
\end{eqnarray}
In the basis (\ref{eq_4}) the components of the pseudo-spin operator have the form
\begin{equation}
\sigma _z  = \left| e \right\rangle \left\langle e \right| - \left| g \right\rangle \left\langle g \right|,\,\,\,\,\sigma _ +   = \left| e \right\rangle \left\langle g \right|\,,\,\,\,\,\,\sigma _ -   = \left| g \right\rangle \left\langle e \right|\,
\label{eq_5}
\end{equation}
while the $Z$ component of the spin $\displaystyle S\ind{z}=\frac{1}{2}\big(|0><0| - |-1><-1|\big)$
and the Hamiltonian of the system $H\ind{S}$ (given by eq. (\ref{eq_2})) reads
\begin{eqnarray}
\displaystyle &&S\ind{z}  = \frac{1}{2}\left[ {\cos \theta \sigma\ind{z}  + \sin \theta \left( {\sigma _ +   + \sigma _ -  } \right)} \right], \nonumber \\
\displaystyle &&H\ind{S}  = \frac{1}{2}\omega \sigma\ind{z} ,\,\,\,\,\,\,\omega  = \left( {\Omega ^2  + \delta ^2 } \right)^{1/2} .
\label{eq_6}
\end{eqnarray}
Taking into account eqs. (\ref{eq_4})-(\ref{eq_6}) we have for the Hamiltonian of a single NV spin interacting with the cantilever
\begin{eqnarray}
\label{eq_7}
\hat H = \frac{1}{2}\omega \sigma ^z  + \omega _r &&\left( {a^ +  a + 1/2} \right) + \frac{\lambda }{2}\left( {a^ +   + a} \right)\times \nonumber \\
&&\left[ {\cos \theta \sigma ^z  + \sin \theta \left( {\sigma ^ +   + \sigma ^ -  } \right)} \right].
\end{eqnarray}
Now we generalize the model given by eq. (\ref{eq_7}) for the system of two
spins (see Fig. \ref{fig_1}). In the absence of an applied magnetic field the system is symmetric and the distance between the spins and the cantilever is equal $h_1=h_2=h$. However, due to the magnetic field induced cantilever deflection, the external magnetic field leads to an asymmetry $h_1\neq h_2$. The imposed asymmetry $\Delta h(B_0)=2|h_1-h_2|/(h_1+h_2)$, $0<\Delta h(B_0)<1$ can be quantified in terms of the amplitude of applied magnetic field $B_0$ and the geometric and material characteristics of the cantilever, as outline din Section \ref{deflection}.

The amplitude of zero point oscillations of the cantilever is of the order of
$a_0=\sqrt{\hbar/(2m\omega\ind{r})}\approx 5\times 10^{-13}m$, where $m\approx 6\times10^{-17}$~kg is the resonator mass and $\omega\ind{r}/(2\pi)\approx 4$~MHz is the first resonance frequency of the cantilever. Details on the cantilever are given in Section~$3$.

This oscillation amplitude is definitely smaller than the asymmetry imposed by the external magnetic field, i.e. $\Delta h(B_0)\gg a_0$. Nevertheless, oscillations of the cantilever produce a varying magnetic field due to the attached magnetic tips, which is proportional to the oscillation amplitude. The key consequence of the asymmetry is that constants of the interaction between the spins and the magnetic tips $\lambda_{1, 2}$ are different for different spins and the Rabi transition frequencies $\Omega_{1, 2}$ are different as well. Therefore, the Hamiltonian of the system of two NV spins interacting with the deformed cantilever reads
\begin{eqnarray}
\displaystyle \hat H =&& \hat H_0  + \hat V, \,\,\,\,  \nonumber \\
\displaystyle \hat H_0  =&& \frac{1}{2}\hbar \omega _1 \sigma _1^z  + \frac{1}{2}\hbar \omega _2 \sigma _2^z  + \hbar \omega _r \left( {a^ +  a + 1/2} \right), \\
\displaystyle \hat V =&& \lambda _1 \left( {a^ +   + a} \right)\frac{1}{2}\hbar \left[ {\cos \theta _1 \sigma _1^z  + \sin \theta _1 \left( {\sigma _1^ +   + \sigma _1^ -  } \right)} \right] + \nonumber \\
&& \lambda _2 \left( {a^ +   + a} \right)\frac{1}{2}\hbar \left[ {\cos \theta _2 \sigma _2^z  + \sin \theta _2 \left( {\sigma _2^ +   + \sigma _2^ -  } \right)} \right]. \nonumber
\label{eq_8}
\end{eqnarray}
Here the term $\hbar \omega\ind{r}(a^+a+1/2)$ describes quantized oscillations
of the cantilever, $\tan \theta_{1,2}=-\frac{\Omega_{1,2}}{\delta}$ and
$\omega_{1,2}=\sqrt{\Omega_{1,2}^2+\delta^2},$ thereby denotes  $\Omega_{1,2}$  the Rabi
frequency of the transition between up and down spin states and $\delta$ is the
detuning between the microwave frequency and the intrinsic frequency of the spins.
In the symmetric case $h_1=h_2=h$, $\omega_1=\omega_2=\omega$ and
$\lambda_1=\lambda_2=\lambda$ the Hamiltonian (\ref{eq_1}) recovers the
previously studied model  \cite{RaCa09,ZhWe10}. The coupling constants between
the spins and the cantilever has the form
$\lambda_{1,2}=g\ind{S}\mu\ind{B}G^{\mathrm{m}}_{1,2}a_0$, where
$g\ind{S}\approx 2$, $\mu\ind{B}$ is the Bohr magneton and
$G^{\mathrm{m}}_{1,2}=\frac{1}{\hat z}|\vec{B}\ind{tip}|$ is the magnetic field
gradient produced by the magnetic tips. Note that if the asymmetry of the deformation
of the cantilever is strong $\Delta h(B_0)\sim 1$, the distance between the
magnetic tip and the nearest to the cantilever adjacent NV spin becomes very
small $z=h_1$ leading to the very large magnetic field gradient
$G^{\mathrm{m}}_{1}=\frac{1}{h_1}|\vec{B}\ind{tip}|$ and to the large
interaction constant $\lambda_1\approx
g\ind{S}\mu\ind{B}G^{\mathrm{m}}_{1,2}a_0/h_1$, while the coupling to the second
spin becomes weak $\lambda_2\approx
g\ind{S}\mu\ind{B}G^{\mathrm{m}}_{1,2}a_0/h_2$ and therefore the relation
$\frac{\lambda_1}{\lambda_2}=\frac{h_2}{h_1}\gg 1$ holds. This means that one
can easily control the interaction between the magnetic tips and the NV spins
simply by tuning the amplitude of the external magnetic field and  thus
controlling the deflection of the cantilever $\Delta h(B_0)$.  In the symmetric
case $z=h_{1,2}=h$, $\Delta h=0$,
$G^{\mathrm{m}}_1=G^{\mathrm{m}}_2=G\ind{m}\approx 10^6$~[T/m] and realistic
values of the parameters are: $h\approx 25$~nm, $\lambda/(2\pi)\approx
0.1$~[MHz], $\omega\ind{r}/(2\pi)\approx 5$~MHz. In what follows the
interaction constant $\lambda$  and the detuning between the cantilever
frequency and the spin splitting $\Delta=\omega\ind{r}-\omega$, $\Delta\approx
2\lambda$ defines the time scale of the problem. A deviation of the values of the constants from the values corresponding to the symmetric case reads
\begin{eqnarray}
\displaystyle \lambda _{1,2}  &&= \lambda \left( {h_{1,2} } \right) = \lambda  \pm \Delta \lambda ,\,\, \Delta _{1,2}  = \Delta \left( {h_{1,2} } \right) = \Delta  \pm \Delta \omega ,\, \nonumber \\
\displaystyle \omega _{1,2}  &&= \sqrt{\left( {\Omega  \pm \Delta \Omega } \right)^2  + \delta ^2 }  = \omega  \pm \Delta \omega ,\,  \Delta \omega  = \frac{{\Omega ^2 }}{\omega }\left( {\Delta h} \right), \\
\displaystyle \Delta \lambda  &&= \lambda \Delta h,\,\, \Delta \Omega  = \Omega \Delta h. \nonumber
\label{eq_9}
\end{eqnarray}
Estimates of the asymmetry parameter $\Delta h(B_0)$ describing a deformation of
the cantilever, for realistic materials and magnetic fields, follow in the next section.

\section{Magnetic-field-induced cantilever deflection}
\label{deflection}

We propose to deposit a magnetic film on the cantilever to explore alternatively
magnetoelastic stress or magnetic-field induced torque to induce a controlled
cantilever deflection due to an external magnetic field $B_0$. In the first case
we exploit the tendency of a film to develop magnetoelastic stress upon magnetization, and this stress induces a curvature of the thin cantilever substrate \cite{sander:99,sander:04,sander:08}. In the second case we exploit the magnetic torque \cite{hopfl:01}  $\vec{T}=\vec{m}\times\vec{B_0}$, where $\vec{m}$ is the total magnetic moment of the film, which is prepared to be oriented along the cantilever length (x-direction), and $\vec{B_0}$ is the applied magnetic field, oriented perpendicularly to the cantilever long axis (z-direction). In this geometry a cantilever deflection along the z-direction results.

For a quantitative determination of the resulting cantilever deflection we specify the cantilever dimensions as follows: length $L=3000$~nm, width  $w=300$~nm, thickness $t_s=30$~nm.  These parameters are well suited for the nano-fabrication of Si cantilevers \cite{saya:02}. They also represent a valid scenario of stress-induced free two-dimensional bending due to the large length-to-width ratio \cite{dahmen:01}.

We assume that the cantilever is fabricated out of Si(100). We use the
corresponding \cite{brantley:73} Young modulus $Y=130$~GPa and Poisson ratio $\nu=0.279$, and density $\rho=2.33\times10^3$ kg/m$^{-3}$. Such a cantilever has a mass of $m=6.29\times10^{-17}$~kg. Its first three resonance frequencies \cite{sarid:97} are calculated from $f_{\rm res}=t_{\rm s}\beta^2(Y/(3\rho))^{0.5}/(4\pi L^2)$  with $\beta=(1.8751, 4.6941, 7.8548)$ as 4.02, 25.2 and 70.6~MHz. This cantilever has a negligible deflection at its end due to its own weight of $3\rho L^4/(2 Y t_{\rm s}^2)=2.4\times10^{-15}$~m, which is more than six orders of magnitude smaller than the field induced deflection, as described next.

\subsubsection{Magnetoelastic-stress-induced cantilever deflection}
Magnetoelastic stress is responsible for the change of length of a bulk sample
upon magnetization, and the resulting strain is known as magnetostriction
\cite{kittel:49,sander:99}. In films, a change of length of the film upon
magnetization is not possible due to the bonding to the substrate, and the film
develops a magnetoelastic stress. This stress induces a curvature of the
substrate, which we exploit to deflect the end of the cantilever. The role of
the external magnetic field is to induce a reorientation of the magnetization of
the film from an in-plane (external magnetic field off) to an out-of-plane
direction (external magnetic field on along $z$-direction). Such a reorientation
of the magnetization direction of the film with thickness $t_{\rm f}$ induces a
corresponding  change of magnetoelastic stress. It induces a deflection of the
free cantilever end given by: \cite{sander:99}

\begin{equation}
defl_{\rm me}=\frac{3 L^2 t_{\rm f} (1+\nu)B_1}{ Y t_{\rm s}^2}
\end{equation}

For a magnetization reversal along the axes of a cubic system the magnetoelastic coupling coefficient $B_1$ enters \cite{sander:99}. For definiteness, we assume a Ni film of thickness 10~nm and we take $B_1$ of bulk Ni, $9.38$~MJ/m$^3$. We note that the effective magnetoelastic coupling in thin films may deviate from its bulk value \cite{sander:99,sander:04,sander:11}, but we stick to the bulk value for this proof of principle case study. With these assumptions we get a deflection of $defl_{\rm me}=27.7$~nm. A Ni film with bulk properties deposited on the top surface of the cantilever has a tendency to contract along the magnetization direction. Thus, the cantilever would be curved upwards for zero external field (magnetization in-plane along $x$), and it would curve downward for magnetic field on (magnetization along $z$). This is irrespective of the sign (along $+z$ or along $-z$) of the external field. For AC magnetic fields ($\omega_{\rm mag}$) the deflection will change with 2 $\omega_{\rm mag}$.

The required deflection can be adjusted by varying the experimental parameters $t_{\rm f}, t_{\rm s}, L$, accordingly. Note that the magnitude of the external magnetic field does not enter. The only requirement is that it is large enough to induce a magnetization reorientation. The effective magnetic anisotropy of the magnetic film should be fairly small to achieve this. It can be tuned by adjusting film thickness, morphology, interface modifications, multilayer structures, and film composition to achieve this goal. The maximum field is given by the requirement $\mu\ind{B}B_0<\hbar\omega_0$, as pointed out above. This gives $B_0<32.7$~mT.

\subsubsection{Magnetic-torque-induced cantilever deflection}
The deposition of a magnetic film on the cantilever opens also the possibility to exploit the torque induced by the external magnetic field along the $z$-direction on the magnetic moments oriented along the $x$-direction of the film to induce a cantilever deflection \cite{sander:98a}. The magnetic torque $\vec{T}=\vec{m}\times\vec{B_0}$ is proportional to both the total magnetic moment  of the film $m$ and the external magnetic field $B_0$. Here, we assume that the magnetic anisotropy of the film is large enough to ensure an in-plane magnetic moment in presence of the out-of-plane field. Thus, the requirement on the magnetic anisotropy of the film differs from that discussed above for the magnetoelastic-stress-induced deflection, as here an effective magnetic anisotropy favoring in-plane magnetization is required.

For definiteness we assume a magnetic moment of $2\mu\ind{B}$ per film atom, and we refer to the atomic volume of bulk Ni, $\rho_{\rm atomic}=1.096\times10^{-29}$~m$^{-3}$ to calculate the total number of Ni atoms in the film. We obtain for the torque-induced deflection \cite{hopfl:01} $defl_{\rm torque}= 4 T L^2/(Y w t^3)$, and from this we find

\begin{equation}
defl_{\rm torque}=\frac{4  t_{\rm f} 2 \mu\ind{B} B_0 L^3}{\rho_{\rm atomic} Y t_{\rm s}^3 }.
\end{equation}

We obtain a deflection of 5.2~nm for the parameters quoted above at a film thickness of $t_{\rm f}=10$~nm for an external field$B_0=10$~mT. Note that here the deflection is proportional to the external magnetic field, which allows a continuous control of the deflection. The magnetic film could be deposited on both sides of the cantilever, which would double the deflection in a given external magnetic field. Larger deflections at a given field are obtained by e.g. increasing the film thickness.

We foresee that the exploitation of the magnetoelastic stress for achieving a
well defined cantilever deflection is experimentally more challenging as
compared to the exploitation of torque. One reason is the required exact tuning
of the magnetic anisotropy. The torque approach is much more robust in that
aspect, as only a sufficiently large energy barrier against out-of-plane
magnetization is needed. Already the shape anisotropy of out-of-plane
magnetization fulfills this requirement \cite{hopfl:01}. Therefore we focus now on the torque-induced deflection.

To quantify the indirect interaction between the NV spins mediated by the
interaction with the magnetic tip we consider again the asymmetry parameter $\Delta h(B_0)=2|h_1-h_2|/(h_1+h_2)$. We calculate the magnetic tip--NV spin distances $h_1,h_2$ as $h_1=h_0-defl$ and $h_2=h_0+defl$, where $h_0$ describes the symmetric case, i.e. the cantilever end is at the center position between both NV spins, which is realized for zero magnetic field for the torque-induced deflection. This gives the asymmetry parameter

\begin{equation}
\Delta h(B_0)=2 \frac{defl_{\rm torque}}{h_0}=\frac{16  t_{\rm f} \mu\ind{B} B_0 L^3}{h_0 \rho_{\rm atomic} Y t_{\rm s}^3 }.
\end{equation}

\section{Effective Hamiltonian: indirect interaction between spins}
\label{interaction}

Interaction of the NV spins with the magnetic tips leads to an indirect
interaction between the NV spins. The Hamiltonian of the indirect interaction
between the NV spins can be evaluated using the Fr\"ohlich method \cite{MiCa05}
\begin{equation}
\displaystyle \hat H\ind{eff}  = \frac{i}
{2}\lambda ^2 \int\limits_{ - \infty }^0 {dt'\left[ {V\left( {t'} \right),V\left( 0 \right)} \right]} ,\,\, \hat V(t) = e^{ - i\hat H_0 t} \hat Ve^{i\hat H_0 t}
\label{eq_10}
\end{equation}
Taking into account eqs. (8) and (\ref{eq_10}) in the rotating wave approximation we deduce
\begin{eqnarray}
\hat H\ind{eff}  =  && - \frac{1}{4} \left\{ \sin ^2 \theta _1 \frac{{\lambda _1^2 }}{{\Delta _1 }}\left( {2n + 1} \right)\sigma _1^z  + \sin ^2 \theta _2 \frac{{\lambda _2^2 }}{{\Delta _2 }}\left( {2n + 1} \right)\sigma _2^z\right\} \nonumber \\
&&-\frac{1}{4} \sin \theta _1 \sin \theta _2 \left( {\frac{{\lambda _1 \lambda _2 }}{{\Delta _1 }} + \frac{{\lambda _1 \lambda _2 }}{{\Delta _2 }}} \right)\left( {\sigma _1^ +  \sigma _2^ -   + \sigma _1^ -  \sigma _2^ +  } \right), \nonumber \\
n = && \langle \langle a^ +  a \rangle \rangle ,\,\,\, \Delta _1  = \omega _r  - \omega _1 ,\,\,\, \Delta _2  = \omega _r  - \omega _2.
\label{eq_11}
\end{eqnarray}
Taking into account that $\sin \theta _{1,2}  \approx \frac{\Omega }{\omega }\left( {1 \pm \frac{{\delta ^2 }}{{\omega ^2 }}\Delta h} \right)$, $\frac{\delta^2}{\omega^2}\ll 1$ and, therefore, $\sin \theta_{1,2}\approx \frac{\Omega}{\omega}=\sin \theta$ we can rewrite the interaction constant in terms of the asymmetry parameter
\begin{eqnarray}
\displaystyle && \lambda _1 \lambda _2 \sin ^2 \theta \frac{{\Delta _1  + \Delta _2 }}{{\Delta _1 \Delta _2 }} \approx \frac{{2\lambda ^2 \Omega ^2 }}
{{\omega ^2 \Delta }} \cdot \frac{{\left( {1 - \left( {\Delta h} \right)^2 } \right)}}{{1 - \left( {\frac{{\Omega ^2 }}{{\omega \Delta }}} \right)^2 \left( {\Delta h} \right)^2 }} = \nonumber \\
&& \frac{{2\lambda ^2 }}{\omega }\left( {\frac{{\Omega ^2 }}{{\omega \Delta }}} \right)\left[ {1 + \left( {\left( {\frac{{\Omega ^2 }}
{{\omega \Delta }}} \right)^2  - 1} \right)\left( {\Delta h} \right)^2 } \right].
\label{eq_12}
\end{eqnarray}
From eq. (\ref{eq_12}) we see that the asymmetry can enhance the interaction between the spins if
\begin{equation}
\displaystyle \frac{{\Omega ^4 }}{{\omega ^2 \Delta ^2 }} > 1.
\label{eq_13}
\end{equation}
In the opposite case,
\begin{equation}
\displaystyle \frac{{\Omega ^4 }}{{\omega ^2 \Delta ^2 }} < 1,
\label{eq_14}
\end{equation}
the asymmetry lowers the interaction strength. Since the variance of parameters is relatively large\cite{RaCa09,ZhWe10} $\frac{\omega\ind{r}}{2 \pi}\approx 1\div 5$~MHz, $\Omega\approx 0.1\div 10$~MHz, $\delta\approx 0.01\div 0.1$~MHz, both cases given by eqs. (\ref{eq_13}) and (\ref{eq_14}) can be realized.

In particular from eq.(\ref{eq_12}) we see that the interaction strength between
spins depends on the asymmetry $\Delta h$ and the dimensionless parameter
$\alpha =\frac{\Omega^{2}}{\omega \Delta}$. Considering standard values of the parameters \cite{RaCa09,ZhWe10} $\omega\approx\Omega,~\lambda=\frac{\Delta}{2}$
from eq. (\ref{eq_12}) for the interaction strength between spins we get:
$\lambda \big[1+\big(\alpha^{2}-1\big)\big(\Delta h\big)^{2}\big]$.
We see that the dependence of the interaction strength between the NV spins on the asymmetry parameter $\Delta h$ is not trivial.
For small values of the parameter $\alpha <1$ the asymmetry arising from the
deformation of cantilever leads to the reduction of interaction strength,
while for $\alpha >1$ the asymmetry increases the interaction strength. In
particular, for the values $\alpha =3$, $\Delta h=0.5$ the interaction between
NV spins is three times larger than the interaction strength in the symmetric case $\sim3\lambda$.
Stronger interaction between spins means larger entanglement. Therefore
controlling  the spin coupling strength we can influence the degree of entanglement
as well. From the physical point of view small values of the parameter
$\alpha\approx \frac{\Omega}{\omega_{r}-\Omega} <1$ corresponds to a large
detuning between oscillation frequency of the cantilever $\omega_{r}$ and the Rabi
frequency $\Omega$s, while for $\alpha >1$ we have the opposite case.

\section{Influence of asymmetry on the entanglement degree}
\label{asymmetry}

In order to study the influence of the asymmetry on the degree of entanglement, we directly solve the Schr\"odinger equation corresponding to the Hamiltonian (\ref{eq_11})
\begin{eqnarray}
\displaystyle && i\frac{{d\left| \psi  \right\rangle }}{{dt}} = \hat H\ind{eff} \left| \psi  \right\rangle , \\
\displaystyle && \left| \psi  \right\rangle = C_1 (t)\left| {\Phi ^ +  } \right\rangle  + C_2 (t)\left| {\Phi ^ -  } \right\rangle  + C_3 (t)\left| {\psi ^ +  } \right\rangle  + C_4 (t)\left| {\psi ^ -  } \right\rangle . \nonumber
\label{eq_15}
\end{eqnarray}
Here $\left| {\Phi ^ \pm  } \right\rangle$ and $\left| {\psi ^ \pm  } \right\rangle$
are Bell states\cite{MiCa05,AmFa08,Woot98}. Taking into account eqs. (\ref{eq_11}), (18) for the resolution coefficients we obtain
\begin{eqnarray}
\displaystyle C_1 (t) =&& C_1 \left( 0 \right)\cos \left( {At} \right) - C_2 \left( 0 \right)\sin (At), \nonumber \\
\displaystyle C_2 (t) =&& C_2 \left( 0 \right)\cos \left( {At} \right) - C_1 \left( 0 \right)\sin (At), \nonumber \\
\displaystyle C_3 (t) =&& \frac{{D^2 C_3 \left( 0 \right) - D\left( {F + B} \right)C_4 \left( 0 \right)}}{{2\left( {F + B} \right)F}}e^{iFt} \\
 &&+ \frac{{D^2 C_3 \left( 0 \right) + D\left( {F - B} \right)C_4 \left( 0 \right)}}{{2\left( {F - B} \right)F}}e^{ - iFt} , \nonumber \\
\displaystyle C_4 (t) =&& \frac{{ - DC_3 \left( 0 \right) + \left( {F + B} \right)C_4 \left( 0 \right)}}{{2F}}e^{iFt} \nonumber \\
 &&+ \frac{{DC_3 \left( 0 \right) + \left( {F - B} \right)C_4 \left( 0 \right)}}{{2F}}e^{ - iFt}, \nonumber
\label{eq_18}
\end{eqnarray}
with the following notations
\begin{eqnarray}
\displaystyle A &&= \frac{{\lambda ^2 }}{{2\omega }}\left( {2n + 1} \right)\frac{{\Omega ^2 }}{{\omega \Delta }}\left[ {1 - \left( {\frac{{\Omega ^2 }}{{\omega \Delta }} - 1} \right)\left( {\Delta h} \right)^2 } \right], \nonumber \\
\displaystyle B &&=  - \frac{{\lambda ^2 \Omega ^2 }}{{2\omega ^2 \Delta }}\left[ {1 - \left( {\frac{{\Omega ^2 }}{{\omega \Delta }} - 1} \right)\left( {\Delta h} \right)^2 } \right], \nonumber \\
\displaystyle D &&=  - \frac{{\lambda ^2 }}{{2\omega }}\left( {2n + 1} \right)\frac{{\Omega ^2 }}{{\omega \Delta }}\left( {\frac{{\Omega ^2 }}{{\omega \Delta }} - 1} \right)\Delta h, \nonumber \\
\displaystyle F &&= \sqrt {B^2  + D^2 }.
\label{eq_20}
\end{eqnarray}
Taking into account eqs. (18), (19) we can quantify the entanglement via the following expression
\begin{equation}
\displaystyle C\left( {\left| {\psi \left( t \right)} \right\rangle } \right) = \left| {\left\langle {\psi ^ *  \left( t \right)} \right|\sigma _y  \otimes \sigma _y \left| {\psi \left( t \right)} \right\rangle } \right|.
\label{eq_21}
\end{equation}
As a result, after straightforward but laborious calculations for concurrence we deduce
\begin{eqnarray}
\displaystyle && \left|C(t)\right| = \left| {\left( {C_1^2(0) - C_2^2(0)} \right)\cos 2At}  \right. \nonumber \\
\displaystyle &&+e^{ - 2iFt} \left( \frac{{\left[ { - DC_3(0) + \left( {F + B} \right)C_4(0)} \right]^2 }}{{4F^2 }}\right) \nonumber \\
\displaystyle &&-e^{ - 2iFt}\left( \frac{{\left[ {D^2 C_3(0) - D\left( {F + B} \right)C_4(0)} \right]^2 }}{{4F^2 \left( {F + B} \right)^2 }} \right) \nonumber \\
\displaystyle &&+e^{2iFt} \left( \frac{{\left[ {DC_3(0) + \left( {F - B} \right)C_4(0)} \right]^2 }}{{4F^2 }} \right) \\
\displaystyle &&-e^{2iFt}\left(\frac{{\left[ {D^2 C_3(0) + D\left( {F - B} \right)C_4(0)} \right]^2 }}{{4F^2 \left( {F - B} \right)^2 }} \right) \nonumber \\
\displaystyle &&+\frac{{\left[ { - DC_3(0) + \left( {F + B} \right)C_4(0)} \right]\left[ {DC_3(0) + \left( {F - B} \right)C_4(0)} \right]}}{{2F^2 }}- \nonumber \\
\displaystyle &&\left. {\frac{{[{DC_3(0) - D\left({F + B}\right)C_4(0)}][{D^2C_3(0) + D\left({F - B} \right)C_4(0)}]}}{{2F^2 \left( {F - B} \right)^2 }}} \right|. \nonumber
\label{eq_22}
\end{eqnarray}

\section{Results}
\label{results}

\begin{figure}[h]
\centering \includegraphics[scale=.5]{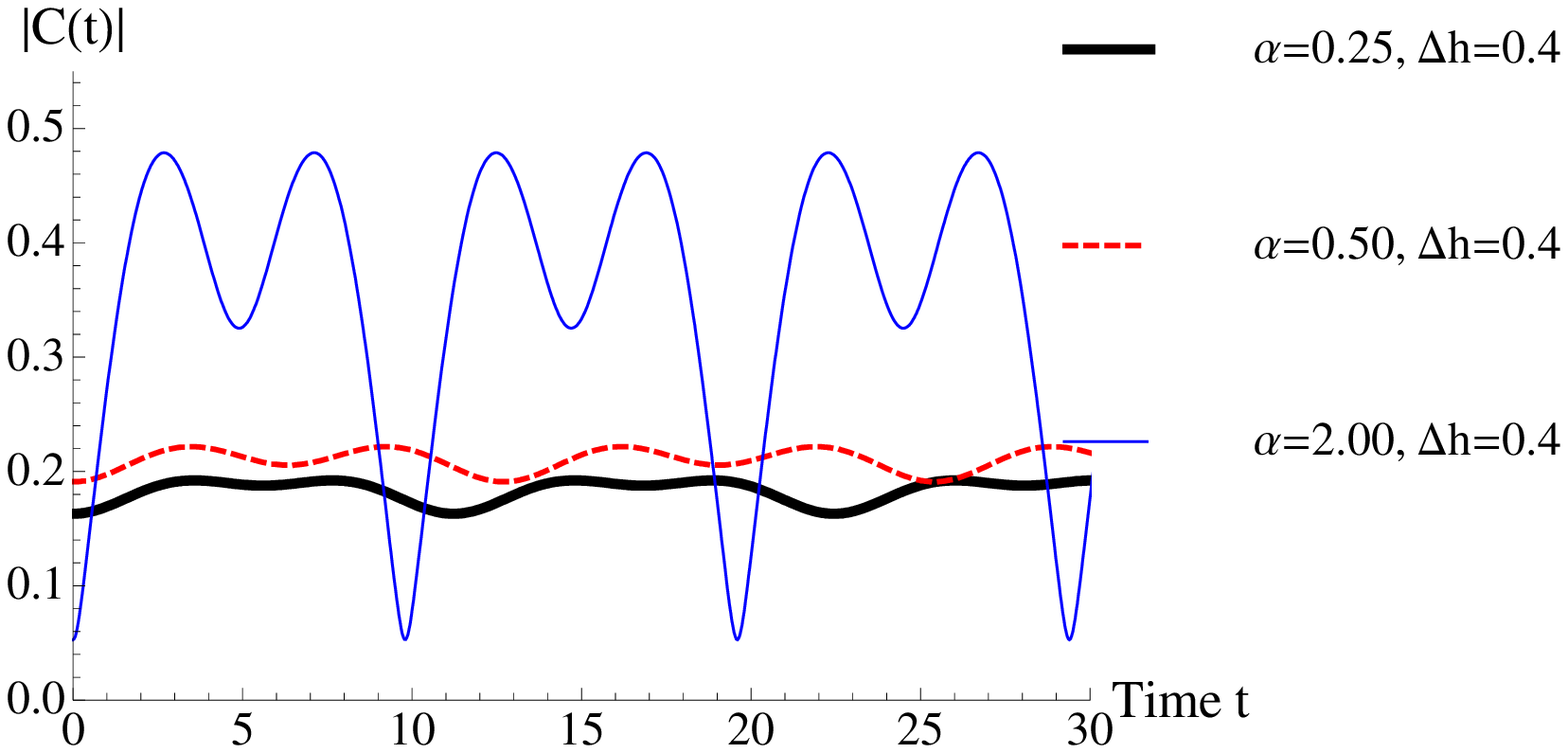}
\caption{Time evolution of the concurrence $|C(t)|$ for the given values of $\alpha$ and $\Delta h$. Other parameters read: $n=1$, $C_1=C_2=C_3=C_4=1/2$. Time-scale corresponds to the microsecond.}
\label{fig_3}
\end{figure}

\begin{figure}[h]
\centering \includegraphics[scale=.5]{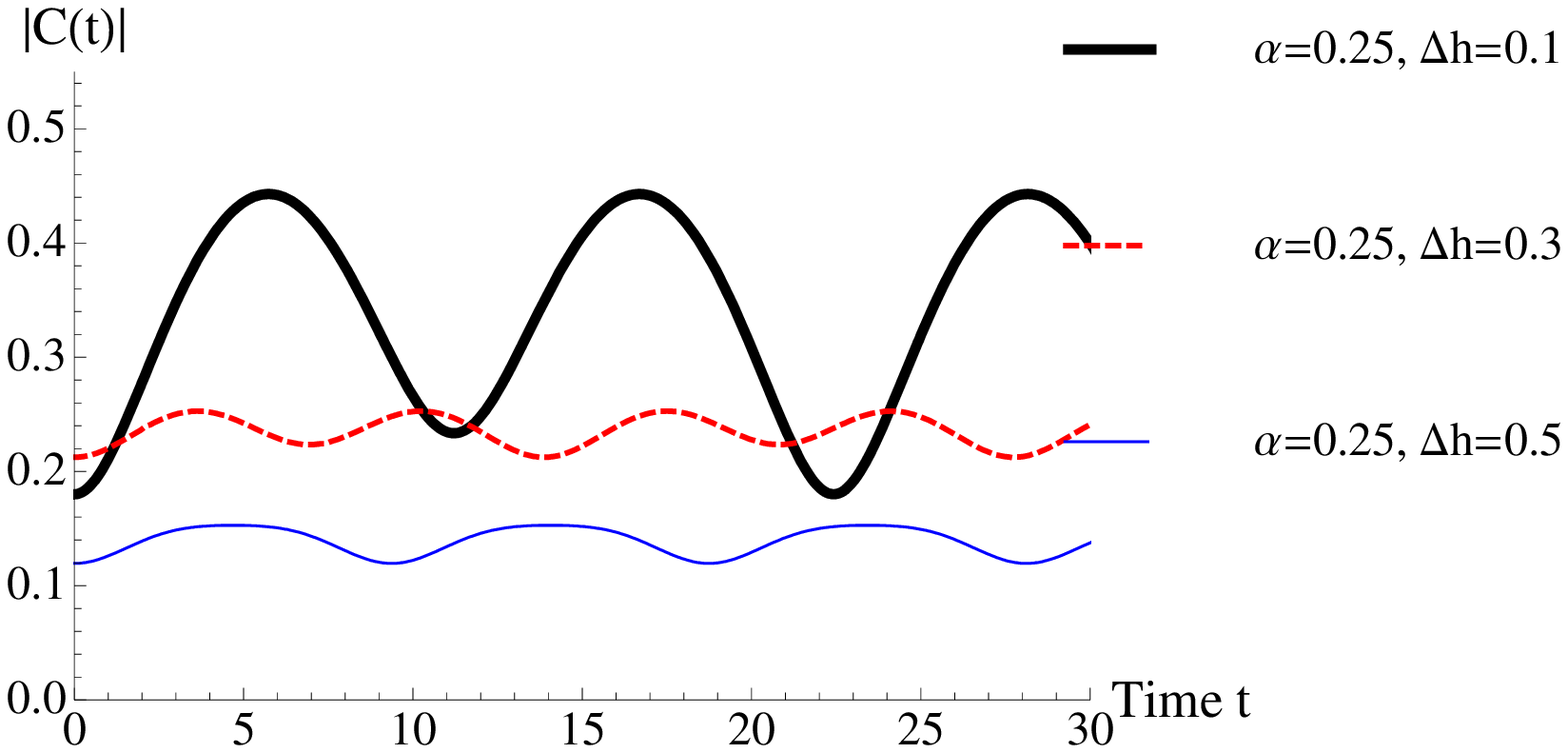}
\caption{Time evolution of the concurrence $|C(t)|$ for the given values of $\alpha$ and $\Delta h$. Other parameters read: $n=1$, $C_1=C_2=C_3=C_4=1/2$. Time-scale corresponds to the microsecond.}
\label{fig_4}
\end{figure}

\begin{figure}[h]
\centering \includegraphics[scale=.5]{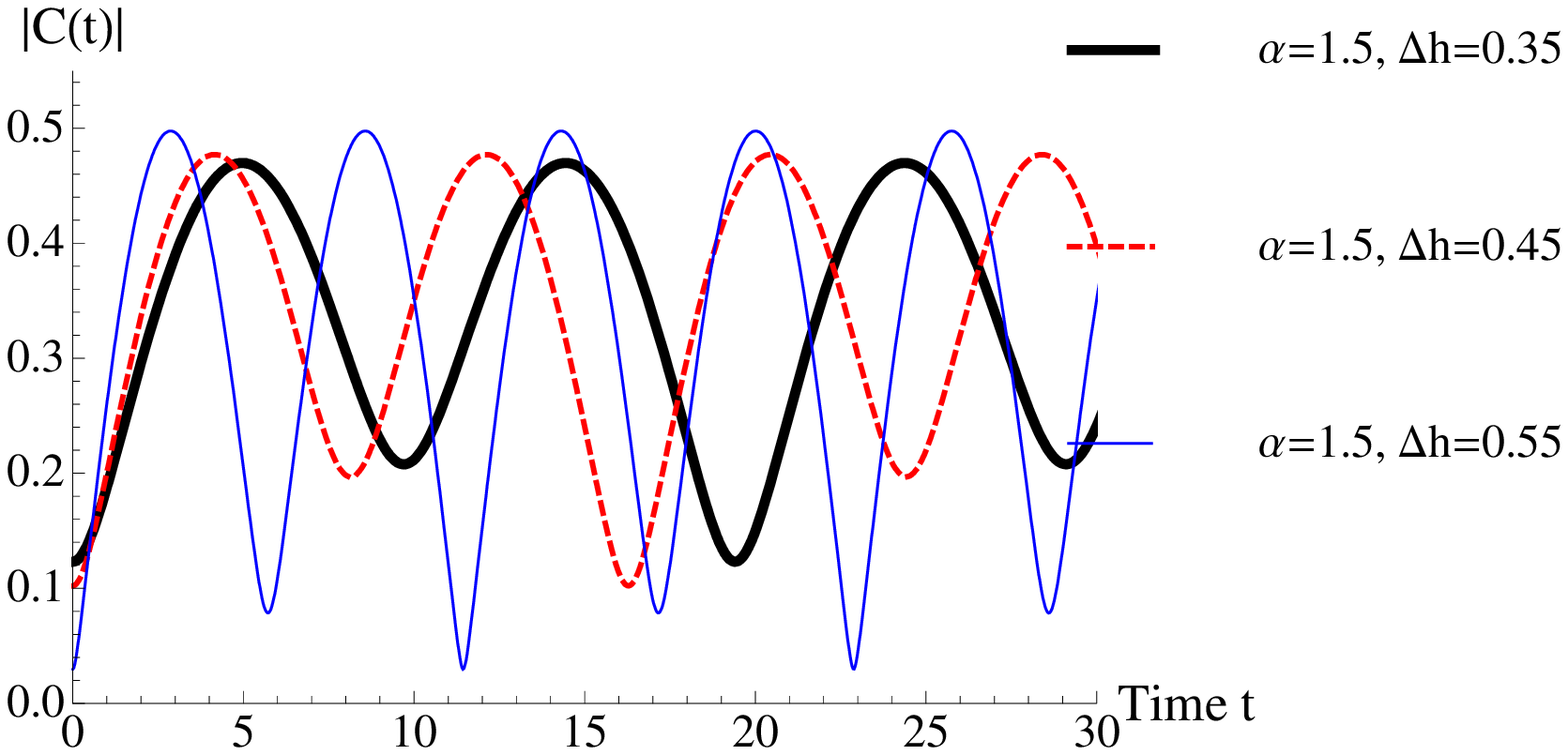}
\caption{Time evolution of the concurrence $|C(t)|$ for the given values of $\alpha$ and $\Delta h$. Other parameters read: $n=1$, $C_1=C_2=C_3=C_4=1/2$. Time-scale corresponds to the microsecond.}
\label{fig_5}
\end{figure}

\begin{figure}[h]
\vspace{-4ex}
\centering \includegraphics[scale=.55]{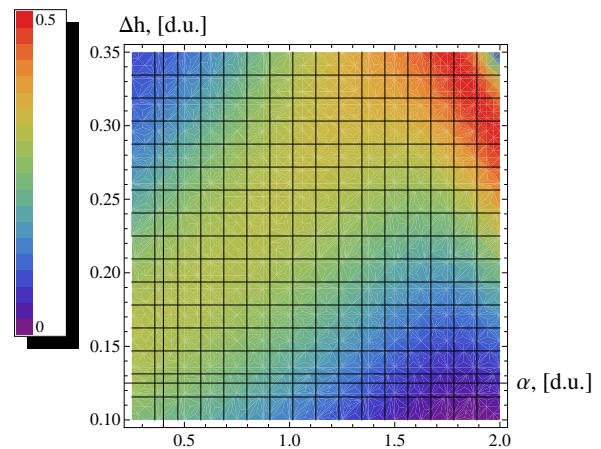}
\caption{Contour plot of the concurrence $|C(t)|$ for $t=13.3$. Other parameters are: $n=1$, $C_1=C_2=C_3=C_4=1/2$.}
\label{fig_6}
\end{figure}

Following eq. (22) the concurrence is plotted in Figs. \ref{fig_3}, \ref{fig_4} and \ref{fig_5} for a variation of two parameters $\alpha$ and $\Delta h$.
The contour plot of $|C(t)|$ is presented in Fig. \ref{fig_6}.

From Fig. \ref{fig_4} we see that for small values of the parameter $\alpha
=0,25$, the increase of the asymmetry $\Delta h$ leads to a smaller
concurrence. The explanation is that for $\alpha <1$ the increase of the
asymmetry $\Delta h$ reduces the coupling strength between spins see eq. (12), while
as we see from Figs. \ref{fig_3}, \ref{fig_5} and \ref{fig_6}, for large $\alpha >1$ concurrence is increased with asymmetry.
In particular, the contour plot of Fig. \ref{fig_6} defines domains of maximal and
minimal concurrence as a function of parameters $\Delta h,~~\alpha$.
For $\alpha <0,5$ concurrence is maximal for small asymmetry $\Delta h<0,25$, while for $\alpha <0,5$ concurrence is maximal for large asymmetry $\Delta h>0,3$.

\section{Conclusions}

One of the main challenges  for  NV spin-based nano-electromechanical resonator
has been the  achievement of a high controlled degree of entanglement and a strong coupling
between NV spins. With this in mind, we proposed in this work
a new type of nano-electromechanical resonator, the functionality of which is
based on the NV spin controlled by an external magnetic field. In particular, we suggest to deposit a thin magnetic Ni film on the
Si(100) cantilever to exploit  alternatively magnetoelastic stress or
magnetic-field induced torque for  inducing a controlled cantilever deflection upon acting
with an external magnetic field. We have shown that, depending on the values of
parameter $\alpha = \frac{\Omega^{4}}{\omega^{2}\Delta^{2}}$, the induced
asymmetry of the cantilever deflection substantially modifies the
characteristics of the system. In particular we demonstrated that is   if $\alpha >1$ the
asymmetry  enhances the strength of the interaction between the NV spins at least three times $\sim 3\lambda$, where $\lambda =100$kHz
is the maximal coupling strength between  NV spins for the symmetric model
reported in \cite{RaCa09}. However, for $\alpha <1$ the asymmetry reduces the interaction strength.
In addition we found that for $ \alpha >1 $
entanglement is maximum for the case of a large asymmetry $\Delta h$, while for
$\alpha <1$ the entanglement is maximal in the small asymmetry case (See Fig. 6).
The values of the parameter $\alpha$ can be changed efficiently via the change of the detuning between the oscillation frequency of the cantilever and the spin splitting frequency $ \Delta=\omega_{r}-\omega $. This can be used as an effective tool for  a practical implementation of our theoretical proposal for controlling the entanglement and the interaction strength between NV spins in  the experiment.

\section{Acknowledgments}The financial support
by the Deutsche Forschungsgemeinschaft (DFG) through SFB 762, contract BE 2161/5-1,
15 Grant No. KO-2235/3 is gratefully acknowledged. This work is partly supported by the National Science Center in Poland as a research project in years 2011 - 2014.

\end{document}